\documentclass[a4paper,english,11 pt,amssymb,nofootinbib,superscriptaddress]{revtex4-2}
\usepackage{lmodern}
\usepackage{lmodern}

\usepackage[T1]{fontenc}
\usepackage[latin9]{inputenc}
\usepackage{babel}
\usepackage{mathrsfs}
\usepackage{amsmath}
\usepackage{amssymb}
\usepackage{esint}
\usepackage[unicode=true,pdfusetitle,
bookmarks=true,bookmarksnumbered=false,bookmarksopen=false,
breaklinks=false,pdfborder={0 0 1},backref=false,colorlinks=false]
{hyperref}

\makeatletter

\@ifundefined{textcolor}{}
{%
	\definecolor{BLACK}{gray}{0}
	\definecolor{WHITE}{gray}{1}
	\definecolor{RED}{rgb}{1,0,0}
	\definecolor{GREEN}{rgb}{0,1,0}
	\definecolor{BLUE}{rgb}{0,0,1}
	\definecolor{CYAN}{cmyk}{1,0,0,0}
	\definecolor{MAGENTA}{cmyk}{0,1,0,0}
	\definecolor{YELLOW}{cmyk}{0,0,1,0}
}

\usepackage{babel}
\usepackage{babel}
\usepackage{babel}
\usepackage{babel}
\usepackage{babel}
\usepackage{babel}
\usepackage{babel}
\usepackage{babel}
\usepackage{babel}
\usepackage{babel}
\usepackage{babel}
\usepackage{babel}
\usepackage{babel}
\usepackage{babel}
\usepackage{babel}
\usepackage{babel}
\usepackage{babel}
\usepackage{graphicx}
\def\b{\begin{equation}}
\def\e{\end{equation}}

\@ifundefined{textcolor}{}{%
	\definecolor{BLACK}{gray}{0}
	\definecolor{WHITE}{gray}{1}
	\definecolor{RED}{rgb}{1,0,0}
	\definecolor{GREEN}{rgb}{0,1,0}
	\definecolor{BLUE}{rgb}{0,0,1}
	\definecolor{CYAN}{cmyk}{1,0,0,0}
	\definecolor{MAGENTA}{cmyk}{0,1,0,0}
	\definecolor{YELLOW}{cmyk}{0,0,1,0}
}

\usepackage{latexsym}\usepackage{bm}

\makeatother

\begin{document}
	\title{Instability of a Kerr-type naked singularity due to light and matter accretion and its shadow}
	
	\author{Aydin Tavlayan}
	
	\email{aydint@metu.edu.tr}
	
	\selectlanguage{english}%
	
	\affiliation{Department of Physics,\\
		Middle East Technical University, 06800 Ankara, Turkey}
	\author{Bayram Tekin}
	
	\email{btekin@metu.edu.tr}
	\affiliation{Department of Physics,\\
		Middle East Technical University, 06800 Ankara, Turkey}

	\selectlanguage{english}%
\begin{abstract}

\noindent We study null and timelike constant radii geodesics in the environment of an over-spinning putative Kerr-type naked singularity. We are particularly interested in two topics: first, the differences of the shadows of the naked rotating singularity and the Kerr black hole; and second, the spinning down effect of the particles falling from the accretion disk. Around the naked singularity, the non-equatorial prograde orbits in the Kerr black hole remain intact up to a critical rotation parameter ($\alpha=\sqrt{6 \sqrt{3}-9}$) and cease to exist above this value [Eur. Phys. J. C \textbf{78}, 879 (2018)]. This has an important consequence in the shadow of the naked singularity if the shadow is registered by an observer on the polar plane or close to it as the shadow cannot be distinguished from that of a Kerr black hole viewed from the same angle considering only the light emanating from the unstable photon orbits.  We show that the timelike retrograde orbits in the equatorial plane immediately (after about an  8\% increase in mass for the case of initial $\alpha=1.5$) reduce the spin parameter of the naked singularity from larger values to $\alpha=1$  at which an event horizon appears. This happens because the retrograde orbits have a larger capture cross-section than the prograde ones. So if a naked singularity happens to have an accretion disk, it will not remain naked for long, an event horizon forms.
\end{abstract}
\maketitle

\section{Introduction}
\noindent Nature has a very efficient way of constraining some physical quantities; it just uses the square roots, for example, the speed of any object is restricted to less than the speed of light as the factor  $\sqrt{1- v^2/c^2}$ appears in relativistic physics. Similarly, in black hole physics, the rotation of a black hole is restricted because the factor $ \sqrt{ 1- \alpha^2}$, with $\alpha$ being the dimensionless rotation parameter given in SI units in terms of the spin $J$ and mass $m$ as $\alpha = c J/(Gm^2)$,  appears in the location of the event horizon. If $ \alpha >1$, there is no event horizon and the black hole becomes a rotating, massive naked singularity, still a solution to vacuum Einstein equations. 

 As the stars or star systems typically have $\alpha >1$ before a black hole is produced, it is clear that the angular momentum of the collapsing matter must be depleted to the values $\alpha <1$ to form a black hole, otherwise a naked singularity is formed.  Even though one can envisage ways to deplete the angular momentum, we still do not know how Nature solves this problem exactly. [For example, for our solar system $\alpha \approx 35$, most of the contribution comes from Jupiter's angular momentum which is located far away from the central mass, the Sun.] Some observed black holes are rotating close to the extreme value $\alpha =1$. With the singularity theorem of Penrose \cite{Penrose1}, a singularity is guaranteed to occur in a gravitational collapse under reasonable assumptions on the energy-momentum tensor of matter. However an event horizon is not guaranteed to form, we only have an expectation dubbed "the cosmic censorship hypothesis" \cite{Penrose2} which states that collapsing matter does not form a naked singularity but it does not state that naked singularities do not exist on their own. Due to this state of affairs, one is necessarily curious about the observable differences in the causal environment of a naked singularity and the Kerr black hole. For example, can a naked singularity mimic the Kerr black hole \cite{Kerr} as far as its shadow \cite{ET} is concerned?  Can accretion of matter to a naked singularity spin down its rotation in such a way that an event horizon forms? 
In this work, we study various null and timelike orbits around a naked singularity and make comparisons with the Kerr black hole; we also give a detailed account of the accretion of matter carrying angular momentum and mass to a naked singularity assuming a thin equatorial accretion disk about it.

The layout of this paper is as follows: In Sec. II we study the null spherical geodesics around the naked singularity. In Sec. III  we plot the shadows of various naked singularities. In Sec. IV we concentrated on the null orbits at the critical inclination angle. In Sec. V we extended the discussion to timelike geodesics. In Sec. VI we give a detailed study of the accretion for both the Kerr black hole and the Kerr-type singularity for thin disks and show the spinning-down effect for the naked singularity due to the matter falling from the unstable interior part of the disk.

\section{Constant radii null geodesics around the Kerr-type naked singularity}
A rotating, massive naked singularity can be obtained from the Kerr metric in the Boyer-Lindquist coordinates $(t,r,\theta,\phi)$ which reads (in the 
$G= c=1$ units) as
\begin{eqnarray}\label{BL}
ds^2=-\Big(1-{{2 m r}\over{\Sigma}}\Big)dt^2-{{4m ar\sin^2\theta}\over{\Sigma}}dt
d\phi+{{\Sigma}\over{\Delta}}dr^2 \nonumber \\
+\Sigma\, d\theta^2+\Big(r^2+a^2+{{2m a^2r\sin^2\theta}\over{\Sigma}}\Big)\sin^2\theta
d\phi^2,
\end{eqnarray}
where $a:=\frac{J}{m}$ is the dimensionfull rotation parameter which we shall take to be $a >m$. The two functions appearing in the metric are given as 
\begin{equation}
\Delta\equiv r^2-2mr+a^2,\hskip 1 cm 
\Sigma\equiv r^2+a^2\cos^2\theta.
\end{equation}
For $a < m$, the larger root of $\Delta =0$  is the event horizon 
located at $r_{\text{H}}=m+(m^2-a^2)^{1/2}$, but in this work, $\Delta \ne 0$ and hence we have a naked rotating singularity. Our first task is to calculate the constant radii null and time-like geodesics in this background. The constant radii null geodesics are particularly important because they are unstable and carry away information about the environment of this strong gravitational region. In practice one computes the shadow of this region as seen by a distant observer. Time-like geodesics are also important as they are traced by massive particles that constitute the accretion disk and change both the mass and the spin of the central object. Studying the evolution of the rotation parameter due to the accretion of matter is our second task.

For the spherical photon orbits, assuming $E \ne 0$ and defining $x:= r/m$, the relevant part of the geodesic equation can be recast as 
\begin{equation}
\Sigma \,{{dx}\over{d\lambda}}=\pm\sqrt{m^2 E^2 {\bf{R}} (x)},
\end{equation}
where $\lambda$ is an affine parameter along the null geodesics and the dimensionless radial function in terms of dimensionless parameters is
\begin{eqnarray}
{\bf{R}}(x):= x^4 + (\alpha^2  - l^2 - q)x^2 
+ 2 x  ( ( \alpha- l)^2 +q)- \alpha^2 q . \label{Rfunction}
\end{eqnarray}
Here $\alpha := a/m$, and $l:=\frac{L_z}{m E}$ where $E$ is the conserved energy  
of the photon corresponding to the time-like Killing vector $\xi_{(t)} = \frac{\partial}{\partial t}$; and $L_z$ is the conserved $z$-component of the angular momentum of the photon related to the $\xi_{(\varphi)} = \frac{\partial}{\partial \varphi}$ Killing vector, while $q:=\frac{{\cal Q}}{ m^2 E^2} $ where ${\cal Q}$ is the Carter's constant related to a symmetric rank two Killing tensor. Explicitly it reads
\begin{equation}
{\cal Q} := p_{\theta}^2+ \cos^2 \theta \left(\frac{L_z^2}{\sin^2 \theta}-a^2 E^2\right), 
\label{Carter1}
\end{equation}
and as a result
\begin{equation}
q= \frac{p_{\theta}^2}{m^2 E^2}+\cos^2 \theta \left(\frac{l^2}{\sin^2 \theta}-\alpha^2\right).
\end{equation}

For constant radii orbits,  ${\cal Q} \ge 0 $, and the bound is satisfied for equatorial orbits  \cite{Wilkins, Teo1, Carter}. There are two conditions on ${\bf{R}}(x)$ for spherical orbits
\begin{equation}
{\bf{R}}(x)=0, \hskip 1 cm   \frac{ d {\bf{R}}(x)}{dx}=0, 
\label{conditions}
\end{equation}
which yield two physically viable equations \cite{Aydin1} \footnote{As discussed in \cite{Teo1}, there is another solution to constant radii conditions which turns out to be unphysical.}:
\begin{eqnarray}
&&l= - \frac{ x^3-3  x^2+\alpha^2 x+\alpha^2 }{\alpha (x-1)}, \label{ang}\\ 
&&q= -\frac{x^3 \left(x^3 -6  x^2 +9 x -4 \alpha^2  \right)}{\alpha^2 (x-1)^2}.\label{Carter}
\end{eqnarray}
Now, we would like to investigate these two equations under various circumstances for  $\alpha>1$.
\subsection{Equatorial null Orbits}
\begin{figure}
	\centering
	\includegraphics[width=0.5\linewidth]{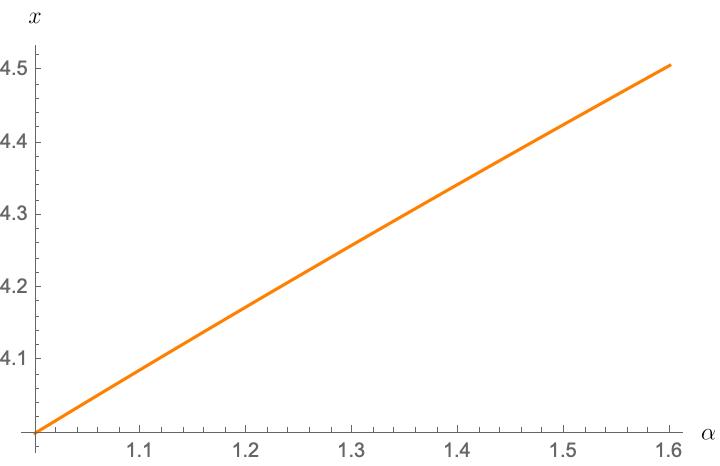}
	\caption{The orbit in the equatorial plane is plotted as a function of the rotation parameter for the interval $1<\alpha<1.6$ using (\ref{retrophoton}). The radius of the orbit increases as the rotation parameter increases, which is an expected result for a retrograde orbit.}
	\label{fig:equatorialretrograde}
\end{figure}

\begin{figure}
	\centering
	\includegraphics[width=0.5\linewidth]{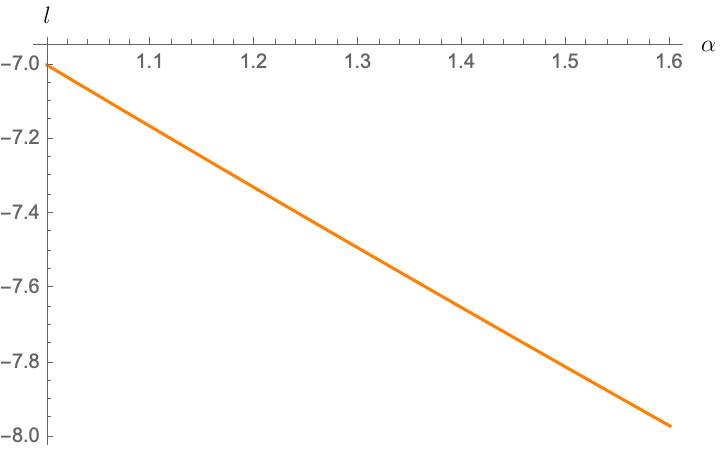}
	\caption{The $l$ value of the equatorial orbit is plotted as a function of the rotation parameter for the interval $1<\alpha<1.6$ using (\ref{ang}). The negative $l$ value confirms that this is a retrograde orbit.}
	\label{fig:equatorialimpact}
\end{figure}

On the equatorial plane, particles with a vanishing $q$ can orbit \cite{Charbulak}. For black holes with a rotation parameter $\alpha<1$, it is known that there are 3 solutions \cite{Aydin1}. One of these lies inside the event horizon while the others are outside the horizon; and the latter correspond to prograde and retrograde orbits. On the other hand, for the naked singularity with a rotation parameter $\alpha>1$, the only real and non-zero solution of (\ref{Carter}) for $q=0$ is \cite{Charbulak}
\begin{eqnarray}
x_-= 2+\left(\alpha +\sqrt{\alpha^2-1}\right)^{2/3}
+\left(\alpha +\sqrt{\alpha^2-1}\right)^{-2/3}
\label{retrophoton}
\end{eqnarray}
which can be seen in Fig. (\ref{fig:equatorialretrograde}). The photons that follow this orbit have a negative $l$ value as shown in Fig. (\ref{fig:equatorialimpact}), and therefore $x_-$ is a retrograde orbit. The equatorial orbits set the {\it innermost }and the {\it outermost} limits of the generic spherical photon orbits. Hence, non-equatorial null orbits can exist only for the interval $0<x<x_-$ in this naked singularity spacetime.

\subsection{Polar Null Orbits}
\begin{figure}
	\centering
	\includegraphics[width=0.5\linewidth]{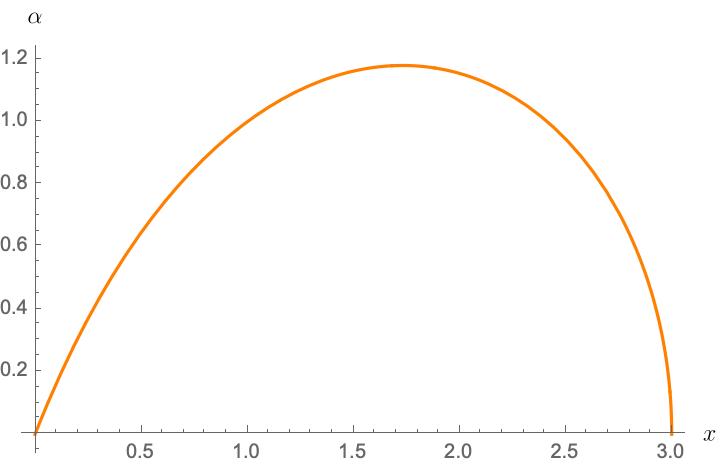}
	\caption{For the vanishing $l$ value, the corresponding rotation parameter as a function of radius is plotted for the interval $0<\alpha<3$ using (\ref{polarrange}). The maximum value of the rotation parameter is $\alpha_{max}=\sqrt{6 \sqrt{3}-9}$. For higher values of rotation parameters, photons cannot reach to the polar plane.}
	\label{fig:rotationparameterlimit}
	\end{figure}

Photons with a vanishing $l$ and a positive $q$ can have orbits on the polar plane. For a black hole with $\alpha<1$, we have already shown that there are 3 polar circular null orbits \cite{Aydin1}. One of them corresponds to an orbit with a negative radius and, therefore physically nonviable.  The second solution lies inside the event horizon. The third solution lies outside the event horizon and corresponds to retrograde orbits.  For the case of a naked singularity, (\ref{ang}) for $l=0$ yields
\begin{equation}
\alpha(x)=\frac{x\sqrt{3-x}}{\sqrt{x+1}},
\label{polarrange}
\end{equation}
which vanishes at $x=0$ and $x=3$, and has a local maximum at $x=\sqrt{3}$ as shown in Fig. (\ref{fig:rotationparameterlimit}). The maximum value of the rotation parameter is $\alpha_{max}=\sqrt{6 \sqrt{3}-9}=1.17996$ as was also found in \cite{Charbulak}. The second derivative ($\frac{ d^2 {\bf{R}}(x)}{dx^2}$ )  for the physically viable orbits shows that while one of them is stable, the other one is unstable.

\begin{figure}
	\centering
	\includegraphics[width=0.5\linewidth]{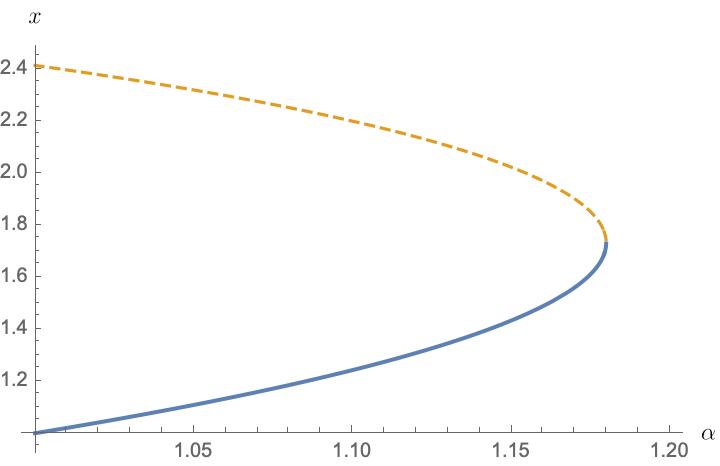}
	\caption{The polar orbits are drawn as a function of the rotation parameter in the interval $1<\alpha<1.2$. See that there are no polar orbits that have $\alpha>\sqrt{6 \sqrt{3}-9}$.}
	\label{fig:polarretropro}
\end{figure}

In Fig. (\ref{fig:polarretropro}), the orbits on the polar plane are plotted as a function of the rotation parameter for the interval $1<\alpha<1.2$. Note that there are no polar orbits for the spacetimes with $\alpha>\sqrt{6 \sqrt{3}-9}$.

Let us note that the value  $\sqrt{6 \sqrt{3}-9}$ is an important limit on the rotation parameter: in the interval $1<\alpha<\sqrt{6 \sqrt{3}-9}$, for generic spherical null orbits, the $l$ value can be positive. This means that prograde, as well as retrograde, orbits are allowed. For a naked singularity with a rotation parameter higher than the maximum rotation parameter, $\alpha>\sqrt{6 \sqrt{3}-9}$, there are only retrograde orbits, prograde orbits simply disappear. This will have an observable consequence in the shadow of the naked singularity as we shall see.

\subsection{Marginally Stable null Orbits}
\begin{figure}
	\centering
	\includegraphics[width=0.5\linewidth]{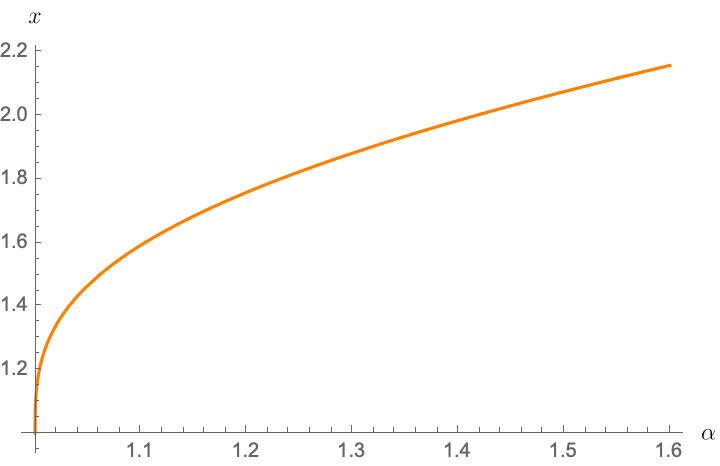}
	\caption{The marginally stable orbit as a function of the rotation parameter is plotted for the interval $1<\alpha<1.6$ using (\ref{marginally}).}
	\label{fig:marginallystable}
\end{figure}

Marginally stable orbits are defined by the condition on the radial function as $\frac{d^2 {\bf {R}}}{dx^2}=0$ augmented with the conditions (\ref{conditions}). On these orbits,  one finds $\alpha(x):=\sqrt{(x-3) x^2+3 x}$, \cite{Charbulak}, or
\begin{equation}
x_M=1+(\alpha^2-1)^{\frac{1}{3}},
\label{marginally}
\end{equation}
which is plotted in Fig.(\ref{fig:marginallystable}). Therefore, orbits with $x<x_{M}$ are stable and orbits with $x>x_{M}$ are unstable. The photons we can observe originate as a result of slight perturbations of the unstable orbits. Hence, stable orbits are not relevant for the shadow imaging. This has an important consequence. The orbits around the naked singularity located at $x=0$ and the orbits around $x=1$ cannot have any visible effects on the image and the shadow.

\section{Shadow of a naked  singularity}
To obtain the shadow, we will use the conventions of \cite{Frolov}, \cite{Patel}. For light rays with parameters $l$ and $q$, we assume that there is an observer located at a point far away from the black hole with coordinates ($r_0,\theta_0,\phi_0$) measuring the directions of these light rays. The coordinate $\phi_0$ can be taken as $0$ using the axial symmetry of the spacetime for simplicity.  $\theta_0$ is called the inclination angle of the observer. At large distances, for light rays one has
\begin{equation}
\frac{d\phi}{dt}\approx\frac{l}{r_0^2 \sin^2\theta_0},
\end{equation}
and
\begin{equation}
\frac{d\theta}{dt}\approx\pm\frac{1}{r_0^2}\sqrt{q+\alpha^2\cos^2\theta_0-l^2\cot^2\theta_0}.
\end{equation}
Observe that the affine parameter is eliminated and the coordinate time $t$ is used. Therefore, on the $2$ dimensional image plane of the observer, we can define the impact parameters, following \cite{Bardeen2} as 
\begin{equation}
X:=-\frac{l}{\sin \theta_0},
\end{equation}
and
\begin{equation}
Y:=\pm\sqrt{q+\alpha^2\cos^2\theta_0-l^2\cot^2\theta_0}.
\end{equation}
Each photon coming from an orbit around the black hole due to a slight perturbation determines a point on the $(X,Y)$ plane of the image taken by the observer. 
\subsection{Two Exemplary Cases}
\subsubsection{Naked singularity with $\alpha=1.1<\alpha_{max}$}
\begin{figure}
	\centering
	\includegraphics[width=0.5\linewidth]{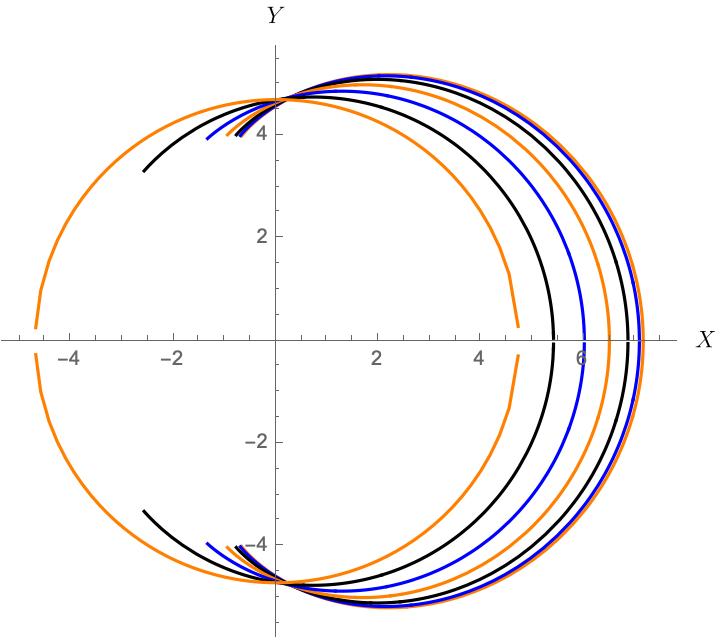}
	\caption{The shadow image of the Kerr-type naked singularity located at the origin, with a rotation parameter $\alpha=1.1$ for observers with different angles $0<\theta_0<\frac{\pi}{2}$. An observer located at the polar plane, $\theta_0=0$, cannot decide whether this is a Kerr black hole or Kerr-type naked singularity considering only the light emanating from the unstable photon orbits. See also\cite{Patel}. }
	\label{fig:differentanglesa11}
\end{figure}

For $\alpha=1.1$, using the results of previous sections, we can find spherical photon orbits with a radius in the interval
\begin{equation}
0<x<4.08808.
\end{equation}
The marginally stable orbit is located at
\begin{equation}x_{M}=1.59439,
\end{equation}
so the unstable spherical photon orbits will be in the interval
\begin{equation}
1.59439<x<4.08808.
\end{equation}
Because we have a rotation parameter that is smaller than $\alpha_{max}=\sqrt{6 \sqrt{3}-9}$, we can expect prograde orbits as well as retrograde orbits. The prograde orbits exist in the interval
\begin{equation}
1.59439<x<2.2,
\end{equation}
while the retrograde orbits are in the interval
\begin{equation}
2.2<x<4.08808.
\end{equation}
The shadow of the black hole is shown in Fig.(\ref{fig:differentanglesa11}). It is important to note that an observer located at the polar plane, $\theta_0=0$, cannot decide whether this is a Kerr black hole or Kerr-type naked singularity.

\subsubsection{Naked singularity with $\alpha_{max}<\alpha=1.5$}
\begin{figure}
	\centering
	\includegraphics[width=0.5\linewidth]{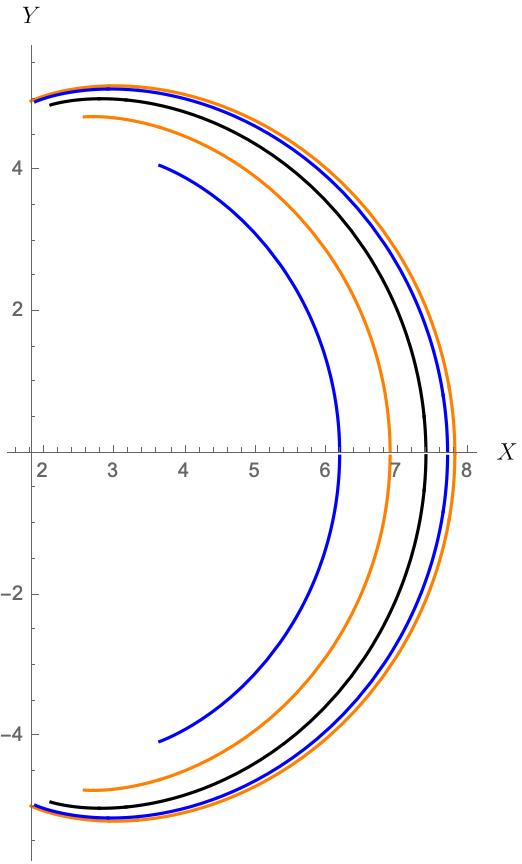}
	\caption{The shadow image of the Kerr-type naked singularity with a rotation parameter $\alpha=1.5$ for observers with different inclination angles $0<\theta_0<\frac{\pi}{2}$. There is no prograde orbit for this spacetime. Therefore, changing the inclination angle only affects the arc shape in the image. See also \cite{Patel}.}
	\label{fig:differentanglesa15}
\end{figure}

For $\alpha=1.5$, we can find spherical photon orbits with a radius in the interval
\begin{equation}
0<x<4.4260.
\end{equation}
The marginally stable orbit is located at
\begin{equation}
x_{M}=2.0772,
\end{equation}
so the unstable spherical orbits will be in the interval
\begin{equation}
2.0772<x<4.4260.
\end{equation}
Because the rotation parameter is greater than $\alpha_{max}=\sqrt{6 \sqrt{3}-9}$, the $l$ value can never change sign, there are no photons that can reach the polar plane, and all the orbits are retrograde. The shadow of the naked singularity is shown in Fig. (\ref{fig:differentanglesa15}).

\section{Critical Inclination Angle for null orbits}
In our previous work, \cite{Aydin1}, by combining the $l$ (\ref{ang}) and $q$ (\ref{Carter}) expressions, we obtained a sextic polynomial and searched its analytical solutions for different conditions. In addition to the known equatorial and polar plane solutions, we found a new family of analytic solutions at the \textit{critical inclination angle}. At this critical inclination angle, the sextic polynomial factors into a quadratic and a quartic part and becomes solvable by radicals. By using the same method, we can find the critical inclination angle solutions for the $\alpha>1$ cases.

\begin{figure}
	\centering
	\includegraphics[width=1\linewidth]{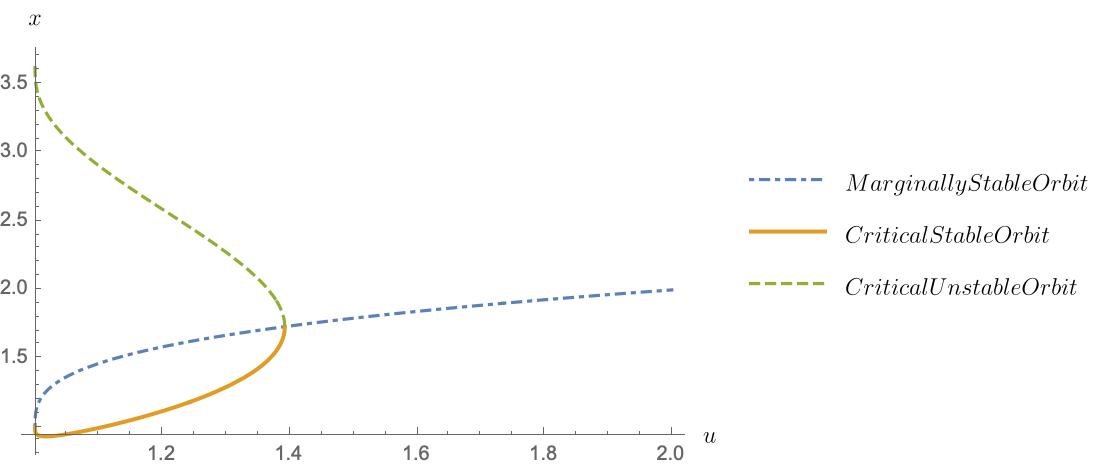}
	\caption{The solution of the sextic polynomial is plotted as a function of the rotation parameter for the interval $1<u<2$.}
	\label{fig:criticalangleorbits}
\end{figure}

\begin{figure}
	\centering
	\includegraphics[width=1\linewidth]{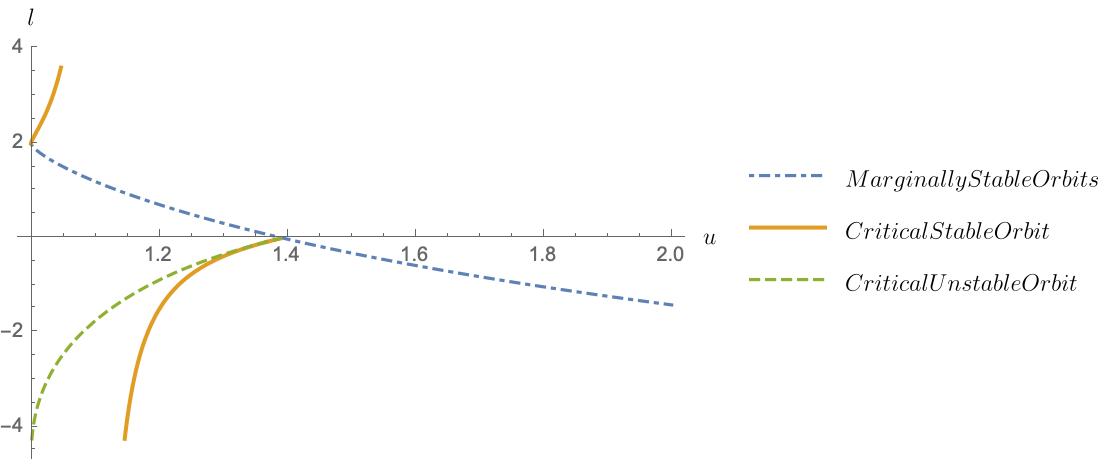}
	\caption{The $l$ value of the critical inclination angle orbit is plotted as a function of the rotation parameter for the interval $1<u<2$.}
	\label{fig:criticalangleimpact}
\end{figure}

\begin{figure}
	\centering
	\includegraphics[width=1\linewidth]{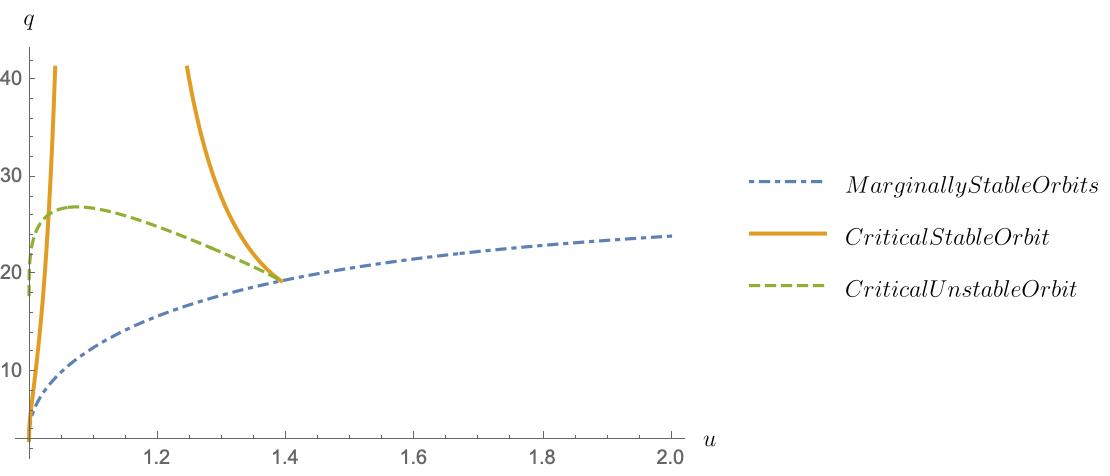}
	\caption{The Carter's constant of the critical inclination angle orbit as a function of the rotation parameter is plotted for the interval $1<u<2$.}
	\label{fig:criticalanglecarter}
\end{figure}

The mentioned sextic polynomial equation is
\begin{eqnarray}
p(x)\equiv x^6-6 x^5+(9+2 \nu u) x^4-4 u x^3-\nu  u (6 -u) x^2 
+2 \nu  u^2 x +\nu  u^2=0,
\label{sextic}
\end{eqnarray}
where we have defined the dimensionless variables
\begin{equation}
 u :=\alpha^2, \hskip 1 cm \nu := \frac{q}{l^2+q} .\ 
\end{equation}
In this section, we will call $u$ to be the rotation parameter.
One must solve this polynomial equation as $x = x(u,\nu)$  for the following intervals:
\begin{equation}
0< x ,\hskip 1 cm   0 \le \nu  \le 1, \hskip 1 cm 1<u. \label{interval}
\end{equation}
To proceed further, it pays to define the following variables which will simplify the final expressions:
\begin{equation}
\nu := \frac{\xi}{u}, \hskip 1 cm u := 1+ w^3,
\end{equation}
with $ 0 \le \xi \le u $ and $w \ge 0 $. Even though a generic radical solution to the sextic (\ref{sextic}) is not possible, it can be shown that it reduces to quadratic times a quartic polynomial at the following critical point for $u>1$:
\begin{eqnarray}
\xi_{\text{cr}}&=&\frac{3 (w+1)^3}{w (w+5)+7}
\label{critical}
\end{eqnarray}
and it becomes solvable. The four real solutions of this sextic polynomial are plotted in Fig. (\ref{fig:criticalangleorbits}).  Two of the solutions are degenerate with a vanishing second derivative and they are marginally stable orbits. One of the remaining solutions has a negative second derivative and the other one has a positive second derivative, therefore they are stable and unstable critical orbits, respectively. The critical stable and unstable orbits are available only for spacetimes with a rotation parameter $u<u_{max}=6 \sqrt{3}-9=1.3923$. The $l$ and $q$ values of these solutions can be seen in Fig. (\ref{fig:criticalangleimpact}) and Fig. (\ref{fig:criticalanglecarter}), respectively. The marginally stable orbits have a turning point and they can be prograde or retrograde. Likewise, the stable orbit can be prograde or retrograde. Yet, the unstable orbit is always retrograde. The Carter's constants are non-zero, as expected. 
\section{Spherical Timelike Orbits around the naked singularity}

Let us now consider a massive particle with mass $\mu$  that moves on a spherical timelike orbit in the vicinity of the naked singularity.
For the spherical timelike orbits, \cite{Aydin2}, the relevant geodesic equation is 
\begin{equation}
\Sigma \,{{dr}\over{d\tau}}=\pm\sqrt{R(r)},
\end{equation}
where the radial function can be rearranged as ${\bf{R}}(x):=R(r)/(m^4 \mu^2)$  which is given as
\begin{eqnarray}
{\bf{R}}(x):= x^2 \left(\alpha^2 \left({\tilde{E}}^2-1\right)-l^2-q\right)-\alpha^2 q
+2 x \left((\alpha {\tilde{E}}-l)^2+q\right)+\left({\tilde{E}}^2-1\right) x^4+2 x^3,\label{Rfunctionmass}
\end{eqnarray}
with $l=\frac{L_z}{m \mu}$, $q=\frac{Q}{m^2 \mu^2}$ and ${\tilde{E}=\frac{E}{\mu}}$.  Here Carter's constant $Q$ was given in (\ref{Carter1}). Note that in contrast to the null geodesics, the energy of the orbit is important, but the mass of the particle is not, hence we scaled out the mass of the particle as well as the mass of the central body.

 Two equations must be satisfied, ${\bf{R}}(x)=0$ and $\frac{ d {\bf{R}}(x)}{dx}=0$, 
for constant radius geodesics. There is a bifurcation of solutions: for $x=1$, one has the following solutions
\begin{equation}
l=\frac{2 \left(\alpha^2+1\right) {\tilde{E}}^2-\alpha^2+1}{2 \alpha {\tilde{E}}},
\end{equation}
and
\begin{equation}
q=\frac{\alpha^2-\left(2 {\tilde{E}}^2+1\right)^2}{4 \alpha^2 {\tilde{E}}^2}.
\end{equation}
On the other hand, for $x \ne 1$, one has 
\begin{eqnarray}
l= \frac{-1}{\alpha^2 (x-1)}\times \left[\alpha {\tilde{E}} (\alpha^2-x^2) 
+\left(x \left(\alpha^3+\alpha (x-2) x\right)^2 \left(\left({\tilde{E}}^2-1\right) x+1\right)\right)^{1/2} \right ],\label{massivel}
\end{eqnarray}
and
\begin{eqnarray}
&&q= \frac{x^2}{\alpha^3 (-1 + x)^2}\times\left[\alpha^3 \left(\left(2 {\tilde{E}}^2-1\right) x+1\right)\right.\label{massiveq}\\
&&\left.+2 {\tilde{E}} \left(x \left(\alpha^3+\alpha (x-2) x\right)^2 \left(\left({\tilde{E}}^2-1\right) x+1\right)\right)^{1/2}\right.\nonumber\\
&&\left.+\alpha x \left(x \left({\tilde{E}}^2 (-((x-4) x+5))+(x-5) x+8\right)-4\right)\right].\nonumber 
\end{eqnarray}
Next, we shall analyze the unit energy solutions for which the equations are more transparent.
\subsection{Unit Energy Timelike Orbits}
\begin{figure}
	\centering
	\includegraphics[width=0.5\linewidth]{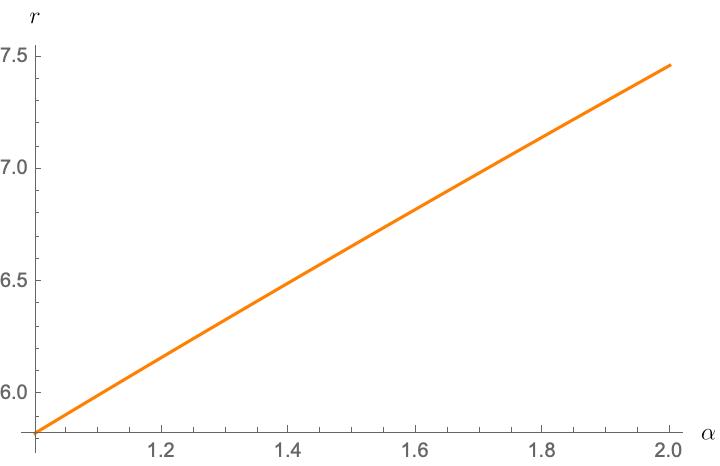}
	\caption{The radius of an equatorial retrograde orbit for a unit energy particle is plotted as a function of the rotation parameter for the interval $1<\alpha<2$ by using (\ref{equatorialbinding1}).}
	\label{fig:massiveunitequatorial}
\end{figure}

\begin{figure}
	\centering
	\includegraphics[width=0.5\linewidth]{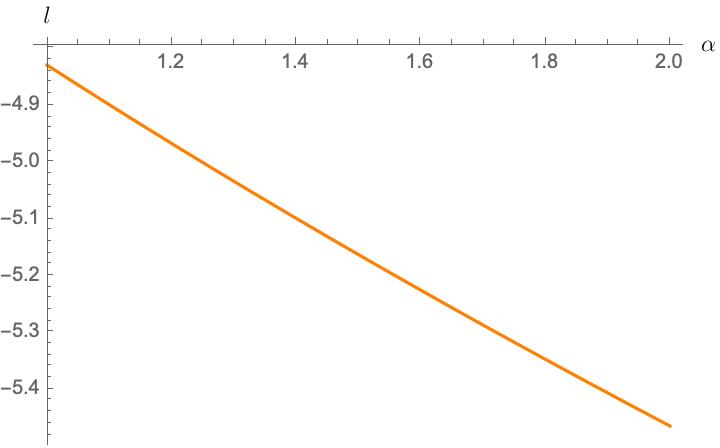}
	\caption{The $l$ value of an equatorial orbit for a unit energy particle is plotted as a function of the rotation parameter for the interval $1<\alpha<2$. The negative $l$ values imply that this is a retrograde orbit.}
	\label{fig:massiveunitequatorialimpact}
\end{figure}

For unit energy particles, ${\tilde{E}}=1$, from (\ref{massivel}) and (\ref{massiveq}) one can obtain for the equatorial plane, $q=0$, 
\begin{equation}
x=2+\alpha \pm 2 \sqrt{\alpha+1},
\label {equatorialbinding1}
\end{equation}
and the plus solution is plotted in Fig. (\ref{fig:massiveunitequatorial}) for the interval $1<\alpha<2$. The $l$ value for this orbit is plotted in Fig. (\ref{fig:massiveunitequatorialimpact}) for the same interval and it shows that this orbit is retrograde.

\begin{figure}
	\centering
	\includegraphics[width=0.5\linewidth]{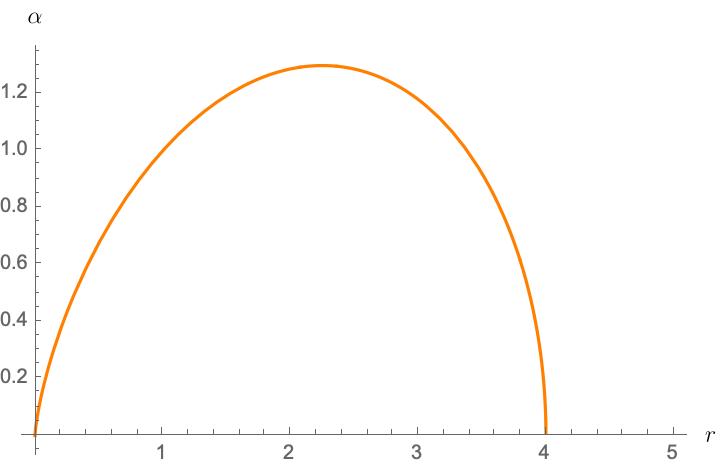}
	\caption{The rotation parameter as a function of the radius of the polar orbits is plotted for the interval $0<x<5$.}
	\label{fig:massiveunitpolarrotation}
\end{figure}

\begin{figure}
	\centering
	\includegraphics[width=1\linewidth]{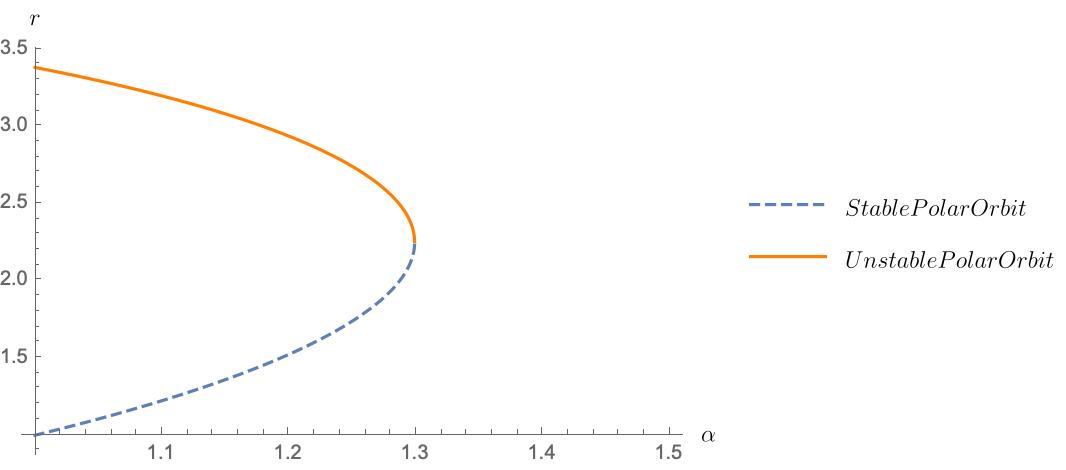}
	\caption{The polar orbits as a function of the rotation parameter are plotted for the interval $1<\alpha<1.5$.}
	\label{fig:massiveunitpolarorbits}
\end{figure}

On the polar plane, $l=0$, we obtain
\begin{equation}
\alpha=\sqrt{2 x^{3/2}-x^2},
\end{equation}
which vanishes at $x=0$ and $x=4$ as can be seen in Fig. (\ref{fig:massiveunitpolarrotation}). This rotation parameter has a maximum at $x_{max}=\frac{9}{4}$ with a value $\alpha_{max}=\frac{3 \sqrt{3}}{4}=1.29904$. For naked singularities with a rotation parameter less than this value, there could be prograde orbits as well as retrograde orbits for generic spherical orbits. It is important to observe that massive particles with unit energy can co-rotate with the black hole for higher values of the rotation parameter than the photons whose maximum rotation parameter is $\alpha_{max}=\sqrt{6 \sqrt{3}-9}$. The spherical timelike orbits of the polar plane can be seen in Fig. (\ref{fig:massiveunitpolarorbits}).
\subsection{Generic Energy orbits on the Equatorial Plane}
\begin{figure}
	\centering
	\includegraphics[width=0.5\linewidth]{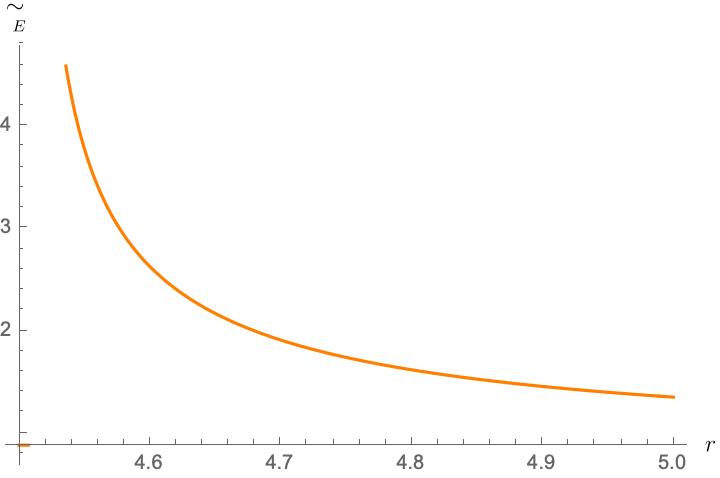}
	\caption{The energy of the particles is plotted as a function of the radius for the interval $4.5<x<5$ around a black hole with a rotation parameter $\alpha=1.6$. Higher energy particles follow orbits with smaller radii.}
	\label{fig:massivegeneralequatorialenergy}
\end{figure}

When we solve the equation (\ref{massiveq}) for $q=0$ concerning the rotation parameter without assuming the unit energy condition, we get
\begin{eqnarray}
\alpha=\left[4x+ x^2 \left(2 {\tilde{E}}^4 x+{\tilde{E}}^2 (5-3 x)+x-4\right)-2 \left(2+\left({\tilde{E}}^2-1\right) x\right) \sqrt{{\tilde{E}}^2 x^3 \left(\left({\tilde{E}}^2-1\right) x+1\right) }\right]^{\frac{1}{2}}.
\end{eqnarray}
Two conclusions can be drawn from this relation. Firstly, for higher rotation parameter values, the radius of the spherical timelike orbit for particles of constant energy on the equatorial plane increases. Secondly, higher energy particles follow closer orbits to the naked singularity with a constant rotation parameter than the lower energy particles. As an example, for $\alpha=1.6$, the energy values as a function of the radius are plotted in Fig. (\ref{fig:massivegeneralequatorialenergy}).

\subsection{Generic Energy orbits on the Polar Plane}
The equation (\ref{massivel}) for  $l=0$ yields
\begin{eqnarray}
\alpha&=&\left(\frac{1}{{\tilde{E}}^2 (x+1)-x}\left[x^2 \left({\tilde{E}}^2 (-(x-1))+x-2\right)+2 \sqrt{{\tilde{E}}^2 x^3 \left(\left({\tilde{E}}^2-1\right) x+1\right)}\right]\right)^{\frac{1}{2}}.
\end{eqnarray}
Then, by using this relation, one can investigate the rotation parameter for different energy values. We have already shown that for a unit energy particle, there is a maximum possible rotation parameter $\alpha_{max}=\frac{3 \sqrt{3}}{4}$. For ${\tilde{E}}=1.5$, one finds $\alpha_{max}=1.21002$ and for ${\tilde{E}}=2$, one finds $\alpha_{max}=1.19495$. In conclusion, one can observe that for the high energy limit (for example ${\tilde{E}}=100$), the maximum value of the rotation parameter approaches the maximum value of the photon case ($\alpha_{max}=\sqrt{6 \sqrt{3}-9}$) as expected.

\section{Accretion into the naked singularity}

\subsection{Review of  the $\alpha<1$ Case}

\subsubsection{Special Circular Null and Timelike Orbits}

Accretion of matter and radiation around both non-rotating and rotating black holes is an extremely important aspect of black hole physics in the strong field region, both from the vantage point of theory and observation. For rotating black holes,  one can consider thin accretion disks around the equatorial plane. One may not be able to solve the whole disk plus the black hole system analytically in an exact form, but one can compute the effects of the accretion disk on the black hole by considering the properties of some special equatorial, circular null, and timelike orbits.  As matter and radiation fall into the black hole, the mass and the spin of the black hole increase as discussed by Bardeen \cite{Bardeen} (without taking into account the radiation) and later by Thorne \cite{Thorne} who considered the effects of the radiation. It turns out the retrograde photon orbits have a larger capture cross-section than that of prograde photon orbits, the latter generically spin up but the former spin down the black hole.
 Before investigating the accretion around a Kerr-type naked singularity, we would like to review the Kerr black hole case with $\alpha<1$.

The radial timelike geodesic equation can be rewritten on the equatorial plane as
\begin{equation}
\,{{dr}\over{d\tau}}=\pm r^{-3/2} \sqrt{R(r)},
\end{equation}
where the radial function can be rearrange to ${\bf{R}}(x):=R(r)/(m^4 \mu^2)$  which is given as
\begin{eqnarray}
{\bf{R}}(x)&:=& \left({\tilde{E}}^2-1\right) x^3+\alpha ^2 \left({\tilde{E}}^2-1\right) x+2 (l-\alpha  {\tilde{E}})^2-l^2 x+2 x^2.
\end{eqnarray}
By using the circularity conditions, ${\bf{R}}(x)=0$ and $\frac{ d {\bf{R}}(x)}{dx}=0$, one can get
\begin{eqnarray}
l_{\pm}&=&\pm\frac{\alpha ^2+x^2 \mp 2 \alpha \sqrt{x}}{\sqrt{\pm2 \alpha  x^{3/2}+(x-3) x^2}},\nonumber\\
{\tilde{E}}_{\pm}&=&\frac{\pm \alpha +(x-2) \sqrt{x}}{\sqrt{{\pm 2 \alpha  x^{3/2}+(x-3) x^2}}},
\end{eqnarray}
where $+$ represents prograde orbits while $-$ represent retrograde orbits. It is important to state that circular orbit solutions exist on the equatorial plane only if $\pm2 \alpha  x^{3/2}+(x-3) x^2 \ge 0 $ and the equality is satisfied only by null orbits which is consistent with the left-hand side of the equations as these quantities represent angular momentum and energy per mass.

The stability of these circular orbits is determined by the second derivative test: $\frac{ d^2 {\bf{R}}(x)}{dx^2}>0$ for unstable orbits, while $\frac{ d^2 {\bf{R}}(x)}{dx^2}<0$  for stable ones. The special case $\frac{ d^2 {\bf{R}}(x)}{dx^2}=0$ represents the innermost stable circular orbit (ISCO) or the marginally stable orbit.  This ISCO condition yields
\begin{equation}
-3 \alpha ^2+x^2 \pm 8 \alpha  \sqrt{x}-6 x=0,
\end{equation}
which has 4 solutions two of which are physically relevant and correspond to prograde and retrograde ISCOs. 

Massive particles with ${\tilde{E}}>1$ follow unstable circular orbits on the equatorial plane and under a slight perturbation, they may escape to infinity or fall into the black hole. Particles with ${\tilde{E}}<1$, under a perturbation, can only fall into the black hole.  The special unstable orbits with ${\tilde{E}}=1$ are aptly called the {\it binding orbits}, and are located at 
\begin{equation}
x_{bind,\pm}=\mp \alpha +2 \sqrt{1 \pm \alpha}+2.
\end{equation}

\subsubsection{Change in the spin of the black hole due to accretion}

Let us assume that there is a subextremal Kerr black hole with a thin accretion disk around its equator, and further assume that there is no gravitational or electromagnetic radiation from the disk. Particles fall into the black hole under slight perturbations from the ISCO. These particles feed the black hole with mass $\delta m = \tilde{E}_{ISCO}$, and angular momentum $\delta J = l_{ISCO}$. Under these assumptions, Bardeen \cite{Bardeen} showed that a black hole that is initially static can be spun up by particles on the accretion disk until it reaches the extremal rotation parameter $\alpha=1$. Later, Thorne \cite{Thorne} calculated the upper limit as $\alpha=0.998$ by considering photons coming out of the accelerated particles. Here it turns out the capture cross-section of the retrograde orbits is larger than that of the prograde orbits and this fact does not allow the subextremal black hole to be spinned up to the extremal value.

\subsection{Accretion around Kerr-type Naked Singularity}

\subsubsection{Special Circular Null and Timelike Orbits}

When the circularity conditions are applied for the radial part of the geodesic equations, ${\bf{R}}(x)=0$ and $\frac{d{\bf{R}}(x)}{dx}=0$, for Kerr-type naked singularity spacetimes, $\alpha>1$, one get
\begin{equation}
{\tilde{E}_\epsilon}=\frac{\left(x-2\right)\sqrt{x}+\epsilon\alpha}{\sqrt{\left(x-3\right) x^2+2\epsilon\alpha x^{3/2}}}, \label{energy45}
\end{equation}
and
\begin{equation}
l_\epsilon=\epsilon\frac{x^2+\alpha^2-2\epsilon\alpha\sqrt{x}}{\sqrt{\left(x-3\right) x^2+2\epsilon\alpha x^{3/2}}}, \label{angular46}
\end{equation}
where $\epsilon=+1$ represents the prograde solutions while $\epsilon=-1$ represents retrograde solutions.
Because the radial part of the geodesic equations has a quadratic dependence on ${\tilde{E}}$ and $l$, there should be a second solution. A straightforward calculation shows that the second solution is
\begin{equation}
{\tilde{E'}}=-{\tilde{E}}, \hspace{0.5 cm} l'=-l.
\end{equation}

As already mentioned, the circular orbits exist only if $(x-3) x^2 \pm 2 \alpha  x^{3/2}\ge 0 $ and equality is satisfied only by null orbits. For spacetimes with $\alpha>1$, the equality is satisfied by a {\it single} orbit which is retrograde. In other words, there is no prograde null orbit on the equatorial plane.

\begin{figure}
	\centering
	\includegraphics[width=0.5\linewidth]{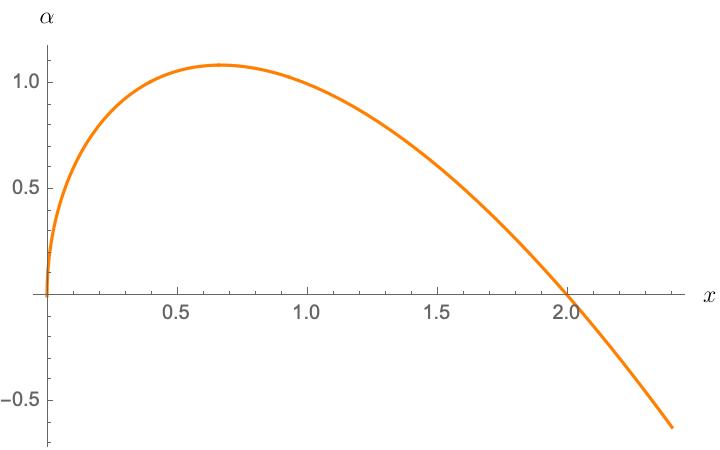}
	\caption{The rotation parameter relation (\ref{rotpar}) is plotted for the interval $0<x<2.4$. The maximum value of the rotation parameter is $\alpha=\sqrt{\frac{32}{27}}$ and this means that it is possible to find orbits with zero or negative energy for the interval $1<\alpha<\sqrt{\frac{32}{27}}$.}
	\label{fig:zeroenergyorbits}
\end{figure}

At this point, let us concentrate on an interesting property of the Kerr-type naked singularity spacetimes, which was first shown in \cite{Felice}. For rotation parameter values which can be given as
\begin{equation}
\alpha\left(x\right)=(2-x) \sqrt{x}, \label{rotpar}
\end{equation}
the particles can rotate in their prograde orbits with zero energy. This is not possible for Kerr spacetimes because these orbits are located inside the event horizon. Yet, because there is no event horizon for naked singularity spacetimes, these orbits are relevant. Even more interestingly, there are orbits with radii smaller than zero energy orbits which are followed by particles with negative energy, \cite{Stuchlik}. By using \ref{rotpar}, one can find that zero and negative energy orbits are possible in the range $1<\alpha< \sqrt{\frac{32}{27}}= 1.08866$ as can be seen in Fig. (\ref{fig:zeroenergyorbits}).

\begin{figure}
	\centering
	\includegraphics[width=0.5\linewidth]{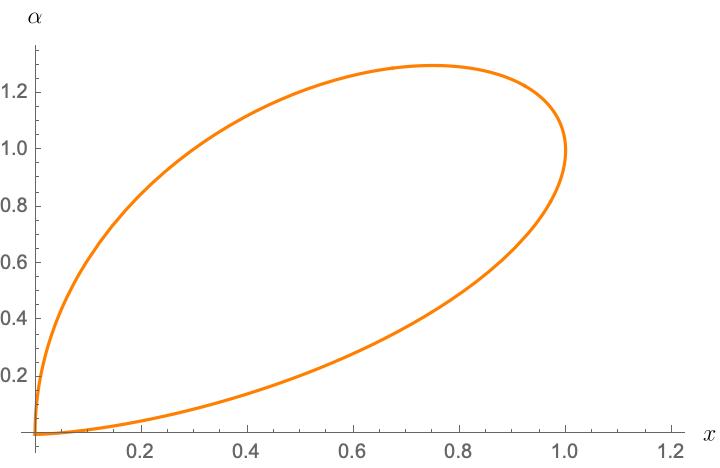}
	\caption{The rotation parameter relation (\ref{rotpar2}) is plotted for the interval $0<x<1.2$. The maximum value of the rotation parameter is $\alpha=\frac{3 \sqrt{3}}{4}$ and this means that it is possible to find orbits with zero or negative angular momentum for the interval $1<\alpha<\frac{3 \sqrt{3}}{4}$.}
	\label{fig:zeroangularmomentumorbits}
\end{figure}

In addition to these zero energy orbits, zero angular momentum orbits or ZAMOs are also possible for Kerr-type naked singularity spacetimes as was shown in \cite{Felice}. For rotation parameter values
\begin{equation}
\alpha\left(x\right)=\sqrt{x} \pm \sqrt{x-x^2},
\label{rotpar2}
\end{equation}
the particles can rotate in their orbits with zero angular momentum. The relation (\ref{rotpar2}) can be seen in Fig. (\ref{fig:zeroangularmomentumorbits}). The maximum of the rotation parameter can be calculated via \ref{rotpar2} and it is $\alpha=\sqrt{\frac{27}{16}}=1.29904$. This means that it is possible to obtain orbits with zero or negative angular momenta for the interval $1<\alpha<\frac{3 \sqrt{3}}{4}$.

The special case, the innermost stable circular orbits (ISCO), correspond to the orbits that satisfy the equation
\begin{equation}
-3 \alpha ^2+x^2 \pm 8 \alpha  \sqrt{x}-6 x=0.
\end{equation}
This condition accepts two solutions which correspond to prograde and retrograde ISCO. For the sake of simplicity, we will not provide the explicit expression here.

The other special case discussed for the Kerr black hole is the binding orbits. There exist prograde and retrograde binding orbits for the Kerr-type naked singularity spacetime which can be given as
\begin{equation}
x_{bind,\pm}=2+\alpha \mp 2 \sqrt{\alpha +1}.
\end{equation}

\begin{figure}
	\centering
	\includegraphics[width=1\linewidth]{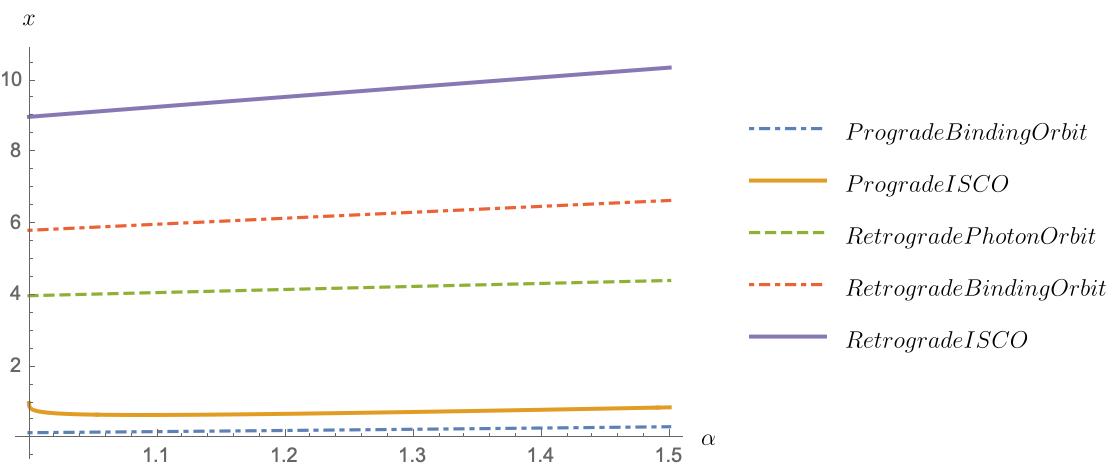}
	\caption{The prograde and retrograde ISCO and binding orbits, and retrograde photon orbit are plotted as a function of the rotation parameter for the interval $1<\alpha<1.5$. Observe that the prograde orbits are closer to the naked singularity than the retrograde orbits which means the latter have a larger capture cross-section. This fact plays an important role in the spinning-down effect. }
	\label{fig:orbitslargera}
\end{figure}

\begin{figure}
	\centering
	\includegraphics[width=0.6\linewidth]{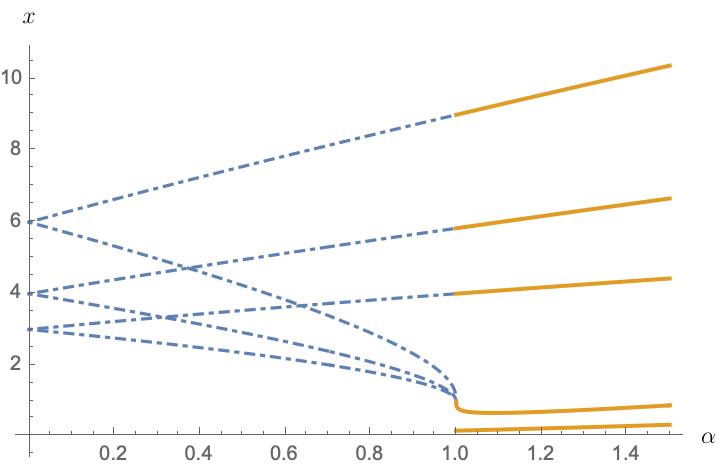}
	\caption{All equatorial orbits are plotted as a function of the rotation parameter for the interval $0<\alpha<1.5$. The dashed lines indicate orbits around the Kerr black hole, while solid lines indicate the orbits around Kerr-type naked singularity. The legends provided for Fig. 18 apply verbatim to this figure.}
	\label{fig:orbitsall}
\end{figure}

The prograde and retrograde ISCO and binding orbits as well as retrograde photon orbits can be seen in Fig. (\ref{fig:orbitslargera}). All orbits on the equatorial plane for both intervals $0<\alpha<1$ and $1<\alpha<1.5$ are plotted in Fig. (\ref{fig:orbitsall}). The retrograde ISCO, binding orbit, and photon orbit continue to exist for the naked singularity spacetimes. The prograde ISCO, binding orbit, and the photon orbit merge at $\alpha=1$. The prograde photon orbit does not exist for $\alpha>1$. The prograde ISCO continues to exist but starts to move away from the singularity with an increasing rotation parameter. For the rotation parameter  $\alpha=\frac{3 \sqrt{3}}{4}$, the particles on the prograde ISCO has $l=0$. The prograde binding orbit also continues to exist but there is a discontinuity at $\alpha=1$ as can be seen in (\ref{fig:orbitsall}).

\subsubsection{Spinning down the Singularity}

An approach similar to the one developed by Bardeen \cite{Bardeen} to investigate the effect of the particles in the accretion disk on the Kerr black hole can be developed for the Kerr-type naked singularity. Let us assume there is a Kerr-type naked singularity with an accretion disk with a negligible thickness on the equatorial plane. Let us also assume gravitational and electromagnetic radiation of this disk is negligible. The gravitational field of the disk itself is much smaller than the gravitational field of the singularity and therefore it is negligible. As a result, it can be assumed that particles on the disk follow circular orbits as already discussed in the previous section.

\begin{figure}
	\centering
	\includegraphics[width=0.7\linewidth]{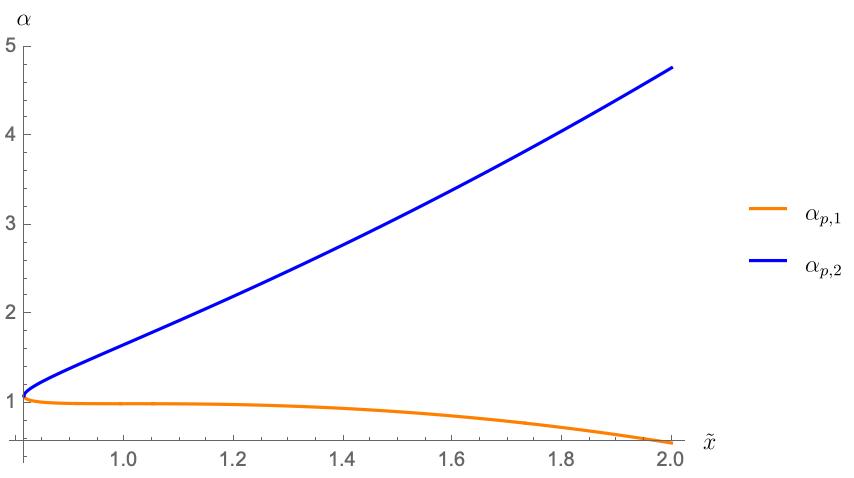}
	\caption{Two rotation parameters, $\alpha_{p,1}$ and $\alpha_{p,2}$, corresponding to prograde orbit are plotted as a function of $\tilde{x}$ for the interval $\sqrt{2/3} \le \tilde{x} \le 2$.}
	\label{fig:alphapro12}
	\end{figure}

As a starting point, the change in the mass and the angular momentum of the Kerr-type naked singularity can be written as
\begin{equation}
\frac{\delta J}{\delta m} = m f\left(\alpha\right)
\end{equation}
to denote their relation with the rotation parameter. Before starting the calculation, to avoid complicated equations, let us define
\begin{equation}
\tilde{x}:=\sqrt{x_{ISCO}}.
\end{equation}
In what follows, we will prove the following fact
\begin{eqnarray}
\delta\left(\ln \tilde{x}\right)&=&-\delta \left(\ln m\right),
\end{eqnarray}
which can also be succinctly written as \footnote{This form was suggested to us by a very conscientious referee who also went over all the computations throughout this work.}
\begin{eqnarray}
\delta\left(m r_{ISCO}\right)&=&0.
\end{eqnarray}
In the previous section, it was shown that, on the innermost stable circular orbit, the condition
\begin{equation}
\tilde{x}^4 - 6\tilde{x}^2 +8 \epsilon \alpha \tilde{x} - 3 \alpha^2 = 0 
\end{equation}
should be satisfied. One can solve this equation with respect to the rotation parameter, $\alpha$. For retrograde orbits for which $\epsilon=-1$, there are two solutions, one of which provides negative rotation parameter values and can be ignored. The physically viable retrograde solution is
\begin{equation}
\alpha_r \left(\tilde{x}\right) = \frac{1}{3} \tilde{x} \left(\sqrt{3 \tilde{x}^2-2}-4\right). \label{aretro}
\end{equation}
Note that for $\tilde{x} \ge 3$, one has $\alpha \ge 1$. 
For prograde orbits for which $\epsilon=+1$, there are two solutions and both of them are physically viable in their corresponding ranges. For the prograde orbit which already appeared in \cite{Bardeen},
\begin{equation}
\alpha_{p,1} \left(\tilde{x}\right) = \frac{1}{3} \tilde{x} \left(4-\sqrt{3 \tilde{x}^2-2}\right),
\end{equation} 
the corresponding range becomes $\sqrt{2/3} \le \tilde{x} \le 1$, and this range ensures that the rotation parameter is $1 \le \alpha \le \sqrt{\frac{32}{27}}$.
The other prograde orbit is
\begin{equation}
\alpha_{p,2} \left(\tilde{x}\right) = \frac{1}{3} \tilde{x} \left(4+\sqrt{3 \tilde{x}^2-2}\right),
\end{equation}
and it exists for $\sqrt{2/3} \le \tilde{x}$ with a rotation parameter value $\sqrt{\frac{32}{27}} \le \alpha$. Both rotation parameter solutions corresponding to prograde orbit can be seen in Fig. (\ref{fig:alphapro12}).
	
At this point, we assume that the infinitesimal changes in the mass and the angular momentum of the naked singularity are equal to those of particles coming from the innermost stable orbit. In other words, we assume that $\delta m = \tilde{E}_{ISCO}$ and $\delta J = l_{ISCO}$. One has
\begin{equation}
f \left(\alpha \right) = \frac{1}{m} \frac{\delta J}{\delta m} = \frac{l_{\pm}}{\tilde{E}_{\pm}}\biggr\rvert_{\tilde{x}}.
\label{fdenk}
\end{equation}

\begin{figure}
	\centering
	\includegraphics[width=0.6\linewidth]{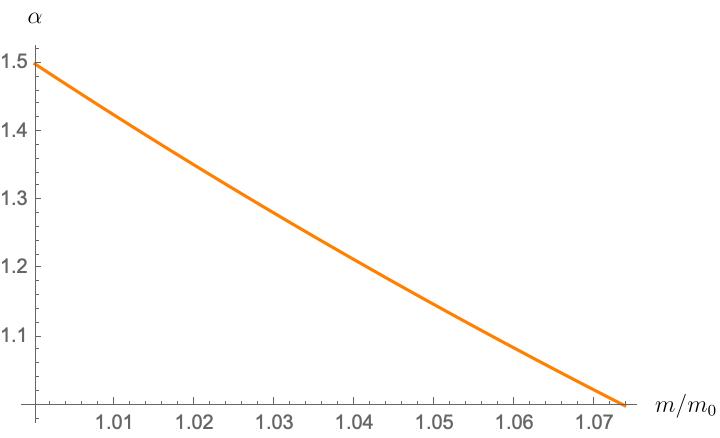}
	\caption{The rotation parameter is plotted as a function of the $m/m_0$ for the interval $ 1 \le \frac{m}{m_0} \le 1.07372$.}
	\label{fig:accretro}
\end{figure}

Let us start calculating the function $f\left( \alpha \right)$ for the retrograde orbits first. 
\begin{eqnarray}
f\left(\alpha\right) &=&  \frac{l_{-}}{\tilde{E}_{-}}\biggr\rvert_{\tilde{x}}=-\frac{2 \tilde{x}}{3} \left(2+\frac{1}{\sqrt{3 \tilde{x}^2 -2}}\right),
\end{eqnarray}
where ${\tilde{E}}_{-}$ was given as \ref{energy45}, $l_{-}$ was given as \ref{angular46} and $\alpha$ was given as \ref{aretro}.
By using the relation $J=\alpha m^2$, one can get
\begin{eqnarray}
f\left(\alpha\right)=\frac{1}{m} \frac{\delta J}{\delta m}
=\frac{\delta \alpha}{\delta \left(\ln m\right)} + 2\alpha. \label{equality}
\end{eqnarray}
With the help of \ref{aretro}, one has
\begin{equation}
\delta \alpha_r = \left(\frac{2}{3} \left(\frac{3 \tilde{x}^2-1}{\sqrt{3 \tilde{x}^2-2}}-2\right)\right) \delta \tilde{x}.
\end{equation}
Hence,
\begin{eqnarray}
f\left(\alpha\right)-2 \alpha &=&-\frac{2 \tilde{x}}{3} \left(2+\frac{1}{\sqrt{3 \tilde{x}^2 -2}}\right)-2 \alpha =-\frac{2 \tilde{x}}{3} \left(-2+\frac{1}{\sqrt{3 \tilde{x}^2 -2}}+\sqrt{3 \tilde{x}^2-2}\right), 
\end{eqnarray}
and
\begin{eqnarray}
\frac{\delta \alpha}{\delta \left(\ln m\right)}&=&\frac{2}{3} \left(\frac{3 \tilde{x}^2-1}{\sqrt{3 \tilde{x}^2-2}}-2\right) \frac{\delta \tilde{x}}{\delta \left(\ln m\right)},
\end{eqnarray}
and by using \ref{equality} one can get
\begin{eqnarray}
\delta\left(\ln \tilde{x}\right)&=&-\delta \left(\ln m\right),
\end{eqnarray}
of which the solution is $\tilde{x}=C/m$ where $C$ is a positive constant and (\ref{aretro}) becomes
\begin{equation}
\alpha_r \left(m\right) = \frac{C}{3m} \left(\sqrt{3 \frac{C^2}{m^2}-2}-4\right). \label{quartic}
\end{equation}
The constant $C$ can be found from the initial conditions
with initial mass $m_0$ and the initial rotation parameter $\alpha_0$. For instance, the retrograde ISCO is located at $x=10.3759$ for $\alpha_0=1.5$; and hence $\tilde{x}=3.22116$. So one finds $C=3.22116 m_0$. After the falling of matter, at a later time, the relation turns into 
\begin{equation}
\tilde{x}=\frac{3.22116 m_0}{m}.
\end{equation}
Therefore, one can write the rotation parameter relation (\ref{aretro}) as a function of the mass of the singularity
\begin{equation}
\alpha\left(m\right) = \frac{1.07372 m_0}{ m} \left(\sqrt{\frac{31.1277 m_0^2}{m^2}-2}-4\right),
\end{equation}
and this relation is valid for $1.07372 \ge \frac{m}{m_0}$. So after accreting a mass of $\delta m=0.07372m_0$ the rotation parameter is reduced to $\alpha_r=1$ from its initial value of $\alpha_r=1.5$. Thus the particles following the retrograde ISCO quickly slow down the rotation of the singularity. This is a rather remarkable result: for example, naked singularity with mass $m_0= 1$ kg and $\alpha_0=1.5$ only requires less than $74$ grams of matter to reduce $\alpha$ to the extremal rotation with an event horizon. The results can be seen in the Fig. (\ref{fig:accretro}). See also the Appendix for an estimate.

\begin{figure}
	\centering
	\includegraphics[width=0.6\linewidth]{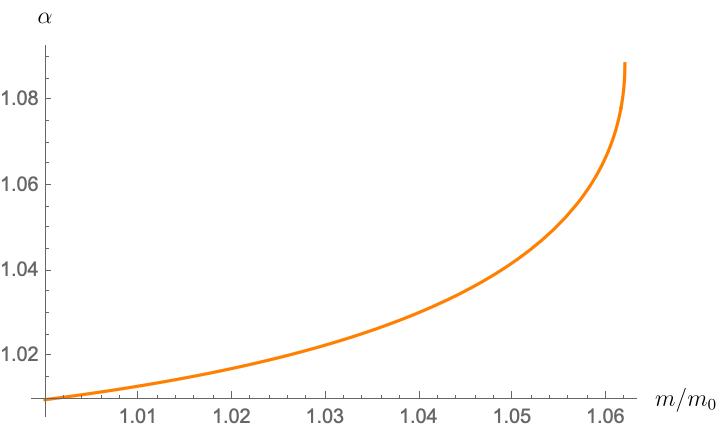}
	\caption{The rotation parameter is plotted with respect to $\frac{m}{m_0}$ for the interval $1 \le \frac{m}{m_0} \le 1.06202$.}
	\label{fig:accpro1}
\end{figure}

Now, we can concentrate on the prograde orbit. For prograde ISCO, from (\ref{fdenk}), one has
\begin{eqnarray}
f\left(\alpha_{p,1}\right)&=&\frac{1}{m} \frac{l_{+}}{\tilde{E}_{+}}\biggr\rvert_{\tilde{x}}=\frac{2 \tilde{x}}{3} \left(2+\frac{1}{\sqrt{3 \tilde{x}^2-2}}\right),
\end{eqnarray} 
and
\begin{eqnarray}
f\left(\alpha_{p,2}\right)&=&\frac{1}{m}\frac{l_{+}}{\tilde{E}_{+}}\biggr\rvert_{\tilde{x}}=\frac{2 \tilde{x}}{3}  \left(2-\frac{1}{\sqrt{3 \tilde{x}^2-2}}\right).
\end{eqnarray}
Let us first study the first prograde solution. Doing the computations as in the retrograde orbit case verbatim, we have
\begin{eqnarray}
f\left(\alpha_{p,1}\right)-2 \alpha_{p,1} &=& \frac{2 \tilde{x}}{3} \left(2+\frac{1}{\sqrt{3 \tilde{x}^2-2}}\right)-2 \alpha_{p,1}=\frac{2 \tilde{x}}{3} \left(\frac{3 \tilde{x}^2-1-2\sqrt{3 \tilde{x}^2-2}}{\sqrt{3 \tilde{x}^2-2}}\right).
\end{eqnarray}                         
One can also find that
\begin{eqnarray}
f\left(\alpha_{p,1}\right)-2 \alpha_{p,1} &=& \frac{\delta \alpha}{\delta \left(\ln m\right)} =-\frac{2}{3}\left(\frac{3\tilde{x}^2-1-2 \sqrt{3 \tilde{x}^2-2}}{\sqrt{3 \tilde{x}^2-2}}\right) \frac{\delta \tilde{x}}{\delta \left(\ln m\right)}. 
\end{eqnarray}
These two equations give
\begin{eqnarray}
\delta \left(\ln \tilde{x}\right)&=&- \delta \left(\ln m\right),
\end{eqnarray}
of which the solution is $\tilde{x}=C/m$ which is valid in the interval $\sqrt{2/3} \le \tilde{x} \le 1$. Let us consider a Kerr-type naked singularity with initial mass $m_0$ and initial rotation parameter $\alpha_0=1.01$. The prograde ISCO is represented by $\alpha_{p,1}$ for this case and it is located at $x=0.75192$ and hence $\tilde{x}=0.867133$. As a result,  $C=0.867133 m_0$. After matter accretion, the relation evolves to
\begin{equation}
\tilde{x} = \frac{0.867133 m_0}{m}.
\end{equation}
As a consequence, the rotation parameter  as a function of mass becomes
\begin{equation}
\alpha_{p,1} \left(m\right)=\frac{0.289044 m_0 }{m}\left(4-\sqrt{\frac{2.25576 m_0^2}{m^2}-2}\right),
\end{equation}
which is valid in the interval $0.867133 \le \frac{m}{m_0} \le 1.06202$. But accretion will not lead to a decrease in mass, so one should restrict this interval to $1 \le \frac{m}{m_0} \le 1.06202$. The evolution of the rotation parameter can be seen in Fig. (\ref{fig:accpro1}).  As can be seen, this prograde orbit tries to increase the rotation parameter value up to $\alpha=\sqrt{\frac{32}{27}}$.

\begin{figure}
	\centering
	\includegraphics[width=0.6\linewidth]{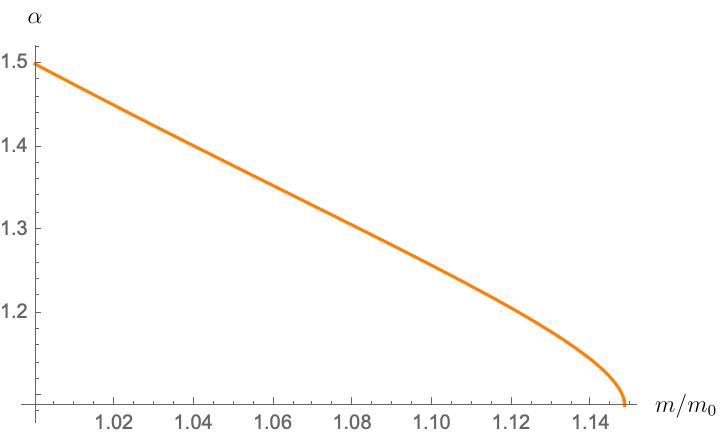}
	\caption{The rotation parameter is plotted as a function of $\frac{m}{m_0}$ for the interval $1< \frac{m}{m_0}<1.14849$.}
	\label{fig:accpro2}
\end{figure}

Now, as a final case, we would like to investigate the second prograde solution. Following similar steps, one arrives at $\tilde{x}=C/m$. Assuming there is a Kerr-type naked singularity with an initial mass $m_0$ and an initial rotation parameter $\alpha_0=1.5$, the prograde ISCO can be found at $x=0.879352$ and hence $\tilde{x}=0.937738$ and $C=0.937738 m_0$. So one has
\begin{equation}
\tilde{x}=\frac{0.937738 m_0}{m}.
\end{equation}
As a consequence, the rotation parameter relation becomes
\begin{equation}
\alpha_{p,2}\left(m\right)=\frac{0.312579 m_0}{m} \left(4+\sqrt{\frac{2.63806 m_0^2}{m^2}-2}\right),
\end{equation}
which is valid for $1.14849 \ge \frac{m}{m_0}$. The evolution of the rotation parameter can be seen in Fig. (\ref{fig:accpro2}). It is interesting to observe that the particles coming from a prograde orbit slow down the rotation. The final rotation parameter value of this process is $\alpha=1.08866$. After that value, the volution is governed by the solution of the second interval.

\begin{figure}
	\centering
	\includegraphics[width=0.6\linewidth]{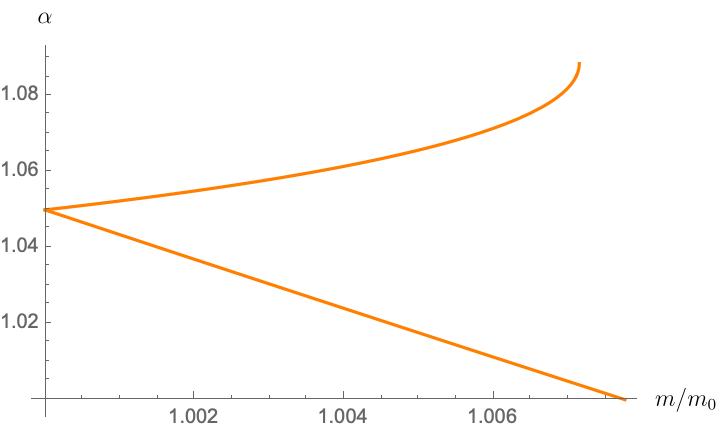}
	\caption{The rotation parameter of a naked singularity with initial mass $m_0$ and rotation parameter $\alpha_0=1.05$ is plotted for both prograde and retrograde orbits as a function of $\frac{m}{m_0}$ for the interval $1< \frac{m}{m_0}<1.00776$. }
	\label{fig:a105both}
\end{figure}

\begin{figure}[h!]
	\centering
	\includegraphics[width=0.6\linewidth]{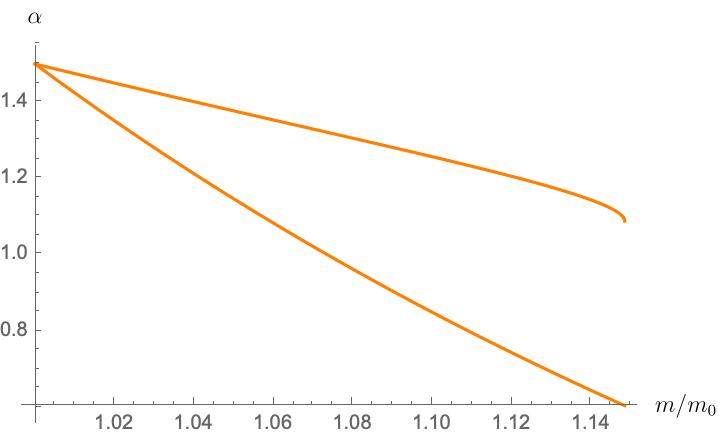}
	\caption{The rotation parameter of a naked singularity with initial mass $m_0$ and rotation parameter $\alpha_0=1.5$ is plotted for both prograde and retrograde orbits as a function of $\frac{m}{m_0}$ for the interval $1< \frac{m}{m_0}<1.14849$. Observe that the retrograde orbits continue to spin down the central object even after the event horizon is formed.}
	\label{fig:a150both}
\end{figure}

To sum up, a Kerr-type naked singularity with an initial mass $m_0$ and an initial rotation parameter $\alpha_0=1.5$ is slowed down by all falling particles. For the retrograde particles, this is easy to understand. Yet, for the prograde particles, it is not: each orbiting particle brings its angular momentum and energy; if the prograde orbit has a higher energy contribution than its angular momentum contribution, then the dimensionless spin is still reduced. This slowing down process continues until $\alpha=\sqrt{\frac{32}{27}}$. Below $\alpha=\sqrt{\frac{32}{27}}$, the particles from the prograde orbit start to spin the singularity up while the particles from the retrograde orbit continue to slow it down as can be seen in Fig. (\ref{fig:a105both}) and Fig. (\ref{fig:a150both}). Due to the larger capture cross-section of the retrograde orbits, in the end, they win over the prograde ones and slow the spin further down until the event horizon appears.

\subsection{Remarks on earlier works}

Let us remark on some of the earlier works about and complementing our discussion in this paper, especially in the context of the time evolution of the parameters of the naked singularity. In  \cite{Stuchlik2}, the optical properties of the "silhouette" and the accretion disc around the Kerr "superspinars" were investigated in depth where constructing the image of the Kerr superspinar was done first, and then, the work focused on the optical properties of the accretion discs around it. While performing the analysis, they provided the Kerr naked singularity and Kerr black hole results for comparison. In another work \cite{Stuchlik3}, a comparison of the effects of counterrotating and corotating orbits around a Kerr superspinar was done. The authors discuss the accumulated mass from both orbits, the conversion times of both orbits and the radiated energy from both orbits. Their conclusion about the conversion time, the time needed to convert a Kerr superspinar into a near-extreme Kerr black hole, is much smaller for counterrotating orbit with respect to corotating orbit. Yet, the energy radiated from the corotating orbit is much higher than the counterrotating orbit. In other words, various Kerr superspinar properties were investigated. For instance, in \cite{Stuchlik4}, radial and vertical epicyclic frequencies of Keplerian motion in the field of Kerr naked singularities were studied; and in \cite{Stuchlik5},  observational properties of Kerr superspinars were demonstrated.

In \cite{Hioki} the question of obtaining the spin parameter and the inclination angle via the shadow of the Kerr black hole and the Kerr-type naked singularity was studied.  For various inclination angles, the appearances of both a Kerr black hole and a rotating singularity are plotted. Our shadow discussion confirms the findings of this earlier work.  Even though our discussion confirms the results of this work, we also discuss the analytically solvable case of the critical inclination angle which was more recently discovered in our earlier work \cite{Aydin1}.

A similar shadow analysis, that was used for the Kerr black hole, Kerr-type naked singularity, and Kerr superspinars, can be performed for many other models. For instance, in \cite{Hou}, the shadows of rotating black holes in the Randall-Sundrum type-II models have been investigated thoroughly. The interesting feature of this work is that they consider not only the near region of the black hole but also the linearized metric in the far region. 

Another interesting paper, \cite{Kumar}, provides a shadow analysis for three rotating regular no-horizon spacetimes, namely, Bardeen, charged Hayward, and nonsingular spacetimes. In that paper, the results were compared with Kerr black hole and Kerr-type naked singularity and they reached a remarkable result: unlike Kerr-type naked singularity spacetimes, which has not a closed shadow, there could be a  closed shadow for no-horizon spacetimes as in the Kerr black hole.

\section{Conclusions}

The Kerr metric is a solution to the vacuum Einstein equations \cite{Kerr} that is assumed to represent the gravitational field of all astrophysical black holes with only two hairs, its mass $m$ and angular momentum $J$. [The Kerr-Newman black hole has the electric charge as the third hair but astrophysical black holes are expected to be neutral.] Such a uniqueness  is so remarkable that it also leads to a rather unique environment which can be observable due to the unstable nature of the photon orbits, among which the constant radii orbits are particularly relevant. In this work, we have studied the Kerr-type naked singularity to understand its shadow and its accretion. We found that a shadow image taken from the polar plane cannot distinguish a naked Kerr-type singularity with a spin parameter up to a maximum value ($\alpha=\sqrt{6 \sqrt{3}-9}$) from a Kerr black hole; while for naked singularities with spins higher than the maximum value, the shadow becomes quite distinct from that of the Kerr black hole. 

We have also studied the timelike orbits that are traced by massive particles around a naked singularity and showed that the rapidly spinning singularity immediately slows down due to the falling matter. Therefore, if a naked singularity is surrounded by a thin accretion disk around its equator, then it slows down and an event horizon is expected to form. In this process, the retrograde orbits play a dominant role as their capture cross-sections are larger than the prograde orbits.  In this present manuscript, we have studied the bare essentials of the shadow and the accretion of Kerr-type naked singularities: this work could be extended in various directions such as studying various other naked singularities in more general gravity theories or studying more realistic accretion disks around such singularities. Without referring to an underlying theory such as General relativity, one can also study the naked singularity version of some phenomenological metrics, such as the Johannsen-Psaltis \cite{JP} metric that modifies the Kerr black hole with some additional parameters.

\section{Appendix: An Estimation of Mass from the Retrograde ISCO}

Using \footnote{This appendix was provided by an anonymous reviewer to whom we thank.} (\ref{quartic}) and the solutions of the quartic equation, one can show that if the initial rotation parameter $\alpha_0$ of the singularity (with initial mass $m_0$) satisfies $\alpha_0>1$, then this parameter is reduced to 1 after the total mass of the particles falling from a retrograde ISCO has reached the value
$$
\bar{m}=\frac{m_0}{3 \sqrt{2}}\left(\sqrt{\beta_{+}+2-\beta_{-}}+\sqrt{\beta_{-}+4-\beta_{+}+\frac{4 \alpha_0 \sqrt{2}}{\sqrt{\beta_{+}+2-\beta_{-}}}}\right),
$$
where
$$
\beta_{ \pm}:=\left(\alpha_0 \pm 1\right)^{2 / 3}\left(\alpha_0 \mp 1\right)^{1 / 3}=\sqrt[3]{\left(\alpha_0^2-1\right)\left(\alpha_0 \pm 1\right)} .
$$

In particular, up to order 2 ,
$$
\frac{\bar{m}}{m_0}-1 \underset{\alpha_0 \rightarrow 1^{+}}{=} \frac{5}{32}\left(\alpha_0-1\right)-\frac{81}{2048}\left(\alpha_0-1\right)^2+o\left(\left(\alpha_0-1\right)^2\right)
$$
thus yielding the equivalent
$$
\frac{\delta m}{m_0} \underset{\alpha_0 \rightarrow 1}{\sim} \frac{5\left(\alpha_0-1\right)}{32}=0.15625\left(\alpha_0-1\right) .
$$

This is a somewhat good approximation as it may be observed that the relative error
$$
\max _{1<\alpha_0<2}\left|\frac{\delta m-5 m_0\left(\alpha_0-1\right) / 32}{\delta m}\right|<11.5 \% .
$$

\end{document}